\documentclass[aps,pra,amsmath,amssymb,10pt,twocolumn,superscriptaddress,notitlepage,floatfix]{revtex4-1}
\usepackage{amsmath,mathtools,amssymb,amsthm}
\usepackage{braket}
\usepackage{graphicx}
\usepackage[lofdepth,lotdepth,caption=false]{subfig}
\usepackage{color}
\usepackage[left=2.5cm,right=2.5cm,top=2.5cm,bottom=2.5cm]{geometry}
\usepackage{array}
\usepackage{booktabs}
\begin{document}
\title{Hypermetallic Polar Molecules for Precision Measurements}
\author{Matthew J. O'Rourke}
\affiliation{
    Division of Chemistry and Chemical Engineering, 
    California Institute of Technology, Pasadena, California 91125, USA
   }
\author{Nicholas R. Hutzler}
\email{hutzler@caltech.edu}
\affiliation{
    Division of Physics, Mathematics, and Astronomy,
    California Institute of Technology, Pasadena, California, 91125, USA
}

\begin{abstract}
Laser cooling is a powerful method to control molecules for applications in precision measurement, as well as quantum information, many-body physics, and fundamental chemistry. However, many optically-active metal centers in valence states which are promising for these applications, especially precision measurement, are difficult to laser cool. In order to extend the control afforded by laser cooling to a wider array of promising atoms, we consider the use of small, hypermetallic molecules that contain multiple metal centers. We provide a detailed analysis of YbCCCa and YbCCAl as prototypical examples with different spin multiplicities, and consider their feasibility for precision measurements making use of the heavy Yb atom. We find that these molecules are linear and feature metal-centered valence electrons, and study the complex hybridization and spin structures that are relevant to photon cycling and laser cooling. Our findings suggest that this hypermetallic approach may be a versatile tool for experimental control of metal species that do not otherwise efficiently cycle photons, and could present a new polyatomic platform for state-of-the-art precision measurements. 

\end{abstract}

\maketitle

\newcommand{\tdm}{\mu_{tr}}

\section{Introduction}

Cold molecules have applications in diverse areas, from quantum information and many-body physics to searches for new physics beyond the Standard Model~\cite{Carr2009,Quemener2012,Safronova2017,Bohn2017}.  All of these applications benefit from (or rely on) the ability to efficiently scatter a large number of photons from a molecule, which allows for laser-cooling and trapping as well as effective quantum state preparation and readout. However, due to their inherently complex internal structures, finding molecules that can efficiently cycle photons is often difficult.  A primary concern is that since there are no selection rules for vibration during an electronic decay, the population of excitations can rapidly become diluted over a large number of internal states.  

Molecules with particular electronic structure and bonding properties avoid this problem~\cite{DiRosa2004,Isaev2016Poly,Kozyryev2016Poly}, which has made laser cooling and trapping of molecules a reality in the last few years~\cite{Tarbutt2019,Shuman2010,Hummon2012,Barry2014,Truppe2017SubDoppler,Anderegg2017}.  Suitable molecules typically have a very simple structure of non-bonding $s$ electrons localized on a metal center~\cite{Ellis2001,Isaev2016Poly,Kozyryev2016Poly}, such as SrF~\cite{Barry2014}, SrOH~\cite{Kozyryev2017SrOH}, CaF~\cite{Truppe2017SubDoppler,Anderegg2017}, YO~\cite{Collopy2018,Hummon2012}, TlF~\cite{Norrgard2017}, YbF~\cite{Smallman2014}, BaF~\cite{Aggarwal2018,Chen2017}, and isoelectronic analogues.  This decouples the electronic and vibrational motion, resulting in highly diagonal Franck-Condon (FC) matrices~\cite{franck1926elementary,condon1926theory}. This property allows these molecules to be laser cooled with a reasonable number of ``repump'' lasers to return excited vibrational states back to the photon cycling process.  Many difficulties remain, including avoiding rotational branching, ``dark states'' that don't scatter photons~\cite{Stuhl2008,Tarbutt2015NJP}, and Renner-Teller effects, but a diagonal FC matrix is a necessary condition for laser cooling methods that rely on spontaneous decay.

Despite the success of this scheme in recent years, molecules useful for precision measurements of fundamental symmetry violations to search for physics beyond the Standard Model pose a number of additional challenges~\cite{Khriplovich1997,Safronova2017}.  First, the molecule must feature a heavy atom that has core-penetrating valence electrons, since sensitivity to CP-violating physics relies on relativistic motion of electrons near the nucleus.  The rapid scaling of this feature with proton number, typically $Z^{2-3}$, effectively restricts the choice of atom to those on the bottom two rows of the periodic table and having valence $s$ or $p$ electrons.  The requirement of efficient photon cycling restricts even further, and generally requires that the atoms have \emph{only} valence $s$ and $p$ electrons.  Thus diatomic species like BaF~\cite{Chen2017}, RaF~\cite{Isaev2010}, TlF~\cite{Norrgard2017}, and YbF~\cite{Smallman2014} are promising candidates for laser cooling and are sensitive to new physics beyond the Standard Model.

However, the requirement of simple electronic structure effectively precludes the advantageous $\Omega-$doublets that arise from electronic orbital angular momentum~\cite{ACME2018,Cairncross2017,Kozyryev2017PolyEDM}.  These nearly-degenerate states of opposite parity can be fully polarized in the lab, leading to ``internal co-magnetometer'' states that are important for rejection of systematic effects.  Species such as ThO~\cite{ACME2018} and HfF$^+$~\cite{Cairncross2017}, which are used in the most sensitive experiments to search for the electron EDM, and other species with experimentally useful $\Omega-$doubled states such as TaN~\cite{Flambaum2014} or WC~\cite{Lee2013}, would be extremely challenging to laser cool using current techniques.

However, polyatomic molecules can offer both photon cycling and fully-polarizable states through their unique vibrational structure~\cite{Kozyryev2017PolyEDM}.  The electronic structure that enables certain diatomic molecules to cycle photons is largely independent of the bonding partner, provided that it has similar valence and ionic nature~\cite{Ellis2001,Isaev2016Poly,Kozyryev2016Poly,Isaev2017RaOH}.  For example, SrOH has similar properties to isoelectronic SrF and was recently laser cooled~\cite{Kozyryev2017SrOH}.  Since sensitivity to CP-violating physics arises from electronic structure at the metal center, molecules like YbOH~\cite{Kozyryev2017PolyEDM,Gaul2018,Denis2019} and RaOH~\cite{Isaev2017RaOH} have sensitivity comparable to their fluoride analogues, but with significant experimental advantages.  Molecules with at least three atoms have nearly-degenerate mechanical modes of opposite parity, such as linear bending modes or symmetric top rotations about the symmetry axis~\cite{Kozyryev2017PolyEDM}.  Molecules of the type MOH, MCCH, MOCH$_3$, and others, where M is a suitable metal such as Yb or Ra, are therefore promising candidates for photon cycling \textit{with} a robust mechanism for rejection of systematic effects.   This could enable efficient state preparation/readout along with the possibility of laser cooling and trapping to achieve long coherence times and perform extremely sensitive searches for CP-violation. For these types of searches, this ability to simultaneously have laser cooling and robust systematic error rejection through parity doublets is unique to polyatomic molecules.

In some sense, the metal atom in these molecules is providing the photon cycling functionality as well as sensitivity to new physics, while the bonding partner is providing the polarization.  A question then arises -- can we attach \emph{multiple} metals with interesting properties to a molecule to realize them simultaneously~\cite{Kozyryev2017PolyEDM}?  For example, a molecule like YbCCCa could provide enhanced scattering rates and advanced co-magnetometry, or a molecule like TaCOCa could be used to laser cool and trap a Ta-containing molecule via photon cycling on the Ca center, thereby enabling access to the advantages of the deformed Ta nucleus for precision measurements~\cite{Flambaum2014}.

In the limit where the two metal centers are infinitely far apart, they will truly be independent and their unique properties can be accessed individually.  However, smaller molecules are more advantageous for practical applications.  In this work, we therefore consider the molecules YbCCCa~\cite{Kozyryev2017PolyEDM} and YbCCAl to explore whether the ``smallest possible'' molecules where the metals do not bond to the same atom can be thought of as two more-or-less independent centers.

Hypermetallic oxides of the form MOM have been studied both experimentally and theoretically \cite{Antonov2011,Ostojic2014,Puri2017}, and recently the mixed hypermetallic BaOCa$^+$ was created and studied in an ion trap~\cite{Puri2017}.  We consider molecules with additional separation between the metals to provide more flexibility in choosing metals and bonding partners.  Additionally, the added distance between the centers should reduce their couplings to each other.

YbCCH~\cite{Loock1997}, CaCCH~\cite{Marr1996}, and AlCCH~\cite{apetrei2007gas} have all been studied spectroscopically, and are linear with low-lying electronic excitations centered on the metal. The species were created via gas-phase chemical reactions of the ablated metal with a reactive gas such as acetylene (HCCH), which suggests a production mechanism for the molecules discussed here.  Yb $(Z=70)$ is sensitive to a range of leptonic and hadronic CP-violating physics while Ca $(Z=20)$ and Al $(Z=13)$ are not, yet they tend to create bonds with higher Frank-Condon factors (FCF).  For example, the 0-0 FCF for the $A\rightsquigarrow X$ transition is $\approx99\%$~\cite{Wall2008} in CaF and $>99.9\%$ in AlF~\cite{DiRosa2004,Wells2011}, compared to $\sim93\%$ for YbF~\cite{Zhuang2011}.  Thus, the hope for these molecules is that the Ca and Al centers will provide better laser cooling than YbOH or YbCCH, but still with similar mass and while maintaining nearly-degenerate states of opposite parity.  As we shall discuss, the Ca and Al metal centers are distinct due to the different sets of possible spin configurations that they permit in the molecules that we consider.

The primary goal of this work is to study the validity of the simple expectation of multiple, quasi-independent cycling centers on these small molecules.  We find that it is indeed the case that the two metal centers can be considered as reasonably independent, and can cycle photons. We also find that the hybridization and spin structure of these exotic molecules plays a critical role in their utility for laser cooling. Our work highlights the potential utility of this hypermetallic approach, and illuminates possibilities for future theoretical and experimental investigations which could explore molecules with heavy metal centers that cannot cycle photons.

\section{Electronic Structure}

A clear trend can be seen in previous works~\cite{Brazier1986,Ellis2001} on molecules such as CaOH~\cite{Isaev2016Poly}, CaNC~\cite{Isaev2016Poly}, SrOH \cite{Kozyryev2017SrOH}, and YbOH~\cite{Kozyryev2017PolyEDM} that metals with alkaline earth-like valence electronic configurations tend to form bonds and hybridized orbitals which are beneficial to creating highly diagonal FC matrices. Thus, a molecule such as YbCCCa, in which both metal centers have an alkaline earth-like valence structure, is a natural place to begin the investigation of small molecules with more than one optically active metal. 

\subsection{YbCCCa}
\label{sec:ybccca_electrons}

\begin{figure*}
  \begin{minipage}[c]{0.65\textwidth}
  \begin{center}
\includegraphics[width=\textwidth]{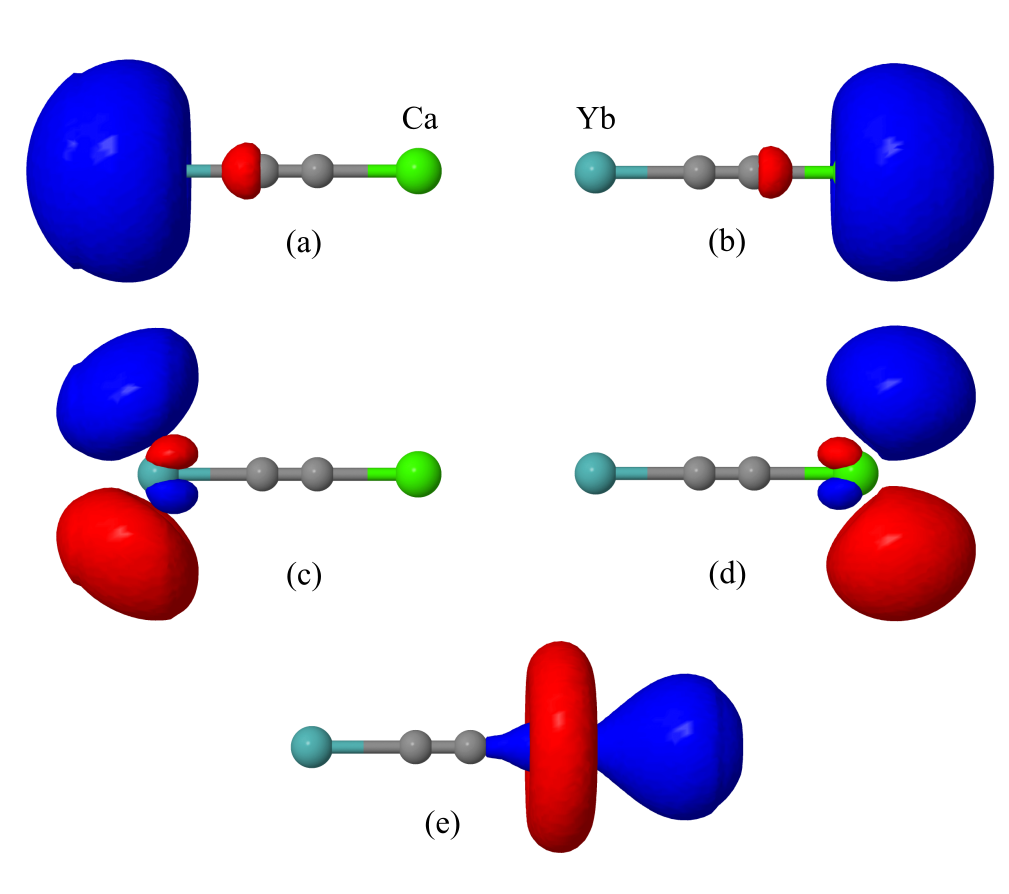}
\end{center}
  \end{minipage}\hfill
  \begin{minipage}[c]{0.30\textwidth}
    \caption{
      Electron and hole orbitals for the ground state and low lying excited states of YbCCCa, computed as the eigenvectors of the difference of density matrices $\rho_{\mathrm{ES}} - \rho_{\mathrm{GS}}$.  \emph{Top:}  The hole orbitals, each of which is occupied in the ground state and unoccupied in the $\Pi$ excited state that is depicted directly below it. (a) shows the unpaired $6s\Sigma$ HOMO-1 on the Yb atom, while (b) shows the unpaired $4s\Sigma$ HOMO on the Ca atom.  \emph{Middle:}  The lowest-lying excited state ($A\Pi$) for the Yb $\Sigma$ electron (c), and the Ca $\Sigma$ electron (d). \emph{Bottom:} The Ca $4s\Sigma \rightarrow 3d\Sigma$ $B$ excited state (e).
    } \label{fig:orbitalsYbCCCa}
  \end{minipage}
\end{figure*}

Using various methodologies from computational quantum chemistry (described in detail in Section \ref{sec:compDetails}), we find that the linear geometry of YbCCCa is lower in energy than various bent and trigonal configurations in both the ground and excited states of interest, which is supported by spectroscopy on similar molecules~\cite{Marr1996,Loock1997}. Additionally, all the excited states of interest lie below the ionization energy. The molecule is open-shell and has the desired bonding pattern, which causes the ground state to have an unpaired $4s\Sigma$ electron on the Ca and an unpaired $6s\Sigma$ electron on the Yb as the highest occupied molecular orbital (HOMO) and HOMO-1, respectively (see Fig. \ref{fig:orbitalsYbCCCa}a-b). The ground state spin structure of these two electrons is characterized by close competition between singlet $X^1\Sigma$ and triplet $X^3\Sigma$ states. This appears to be a significant piece of evidence that the cycling electrons on the two metal centers are highly independent, because in the limit where they are truly independent we expect the singlet and triplet states to be exactly degenerate. The computed splitting between these two states in YbCCCa is approximately $10^{-3}$ eV, with the singlet lying lower than the triplet. This value is considered quite small to resolve with high certainty using standard quantum chemistry methods, but even if one assumes a large error of $\pm 50 \%$ it is still easily resolved experimentally as it corresponds to a frequency on the order of $\sim 100$ GHz.  Fortunately, neither the precise size of this splitting nor the ordering of the states are critical to our conclusions, provided that the gap between the states is larger than typical radiative widths of $\sim 10$~MHz.

The structure of the lowest lying excited states primarily consists of $4s\Sigma \to 4p\Pi$ transitions on the Ca atom (which we will call the Ca $A$ state),  $6s\Sigma \to 6p\Pi$ on the Yb atom, (which we will call the Yb $A$ state), and $4s\Sigma \to 3d\Sigma$ on the Ca atom (which we will call the Ca $B$ state), which are the potential laser cooling transitions. We use this nomenclature because these transitions are analogous to the $X^2\Sigma\rightarrow A^2\Pi$ and $X^2\Sigma\to B^2\Sigma$ transitions in the single-center molecules~\cite{Ellis2001}. In both the $X\Sigma$ ground state and $A\Pi$ excited states, we observe similar molecular orbital hybridization to the molecules studied in Ref. \cite{Isaev2016Poly}, for which those authors gave a detailed discussion about the likely advantages of this structure for diagonal FC matrices. The molecular orbitals of the promoted electrons and their corresponding holes are shown in Figure \ref{fig:orbitalsYbCCCa} for all of these excited states. These orbitals are constructed as the eigenvectors of the difference of the density matrices for the exited state and the ground state, $\rho_{\mathrm{ES}} - \rho_{\mathrm{GS}}$, which is dominated by two nonzero eigenvalues ($\approx \{+1,-1\}$) corresponding to the electron and hole, respectively. The equilibrium bond lengths for the ground and exited states of interest, as well as the Yb--C and C--Ca bond energies and the permanent dipole moment of the molecule are given in Table \ref{tab:bondlengthYbCCCa}.

\begin{table}[t]
    \centering
    \begin{tabular}{ccc@{\hskip 4mm}c@{\hskip 4mm}ccc}
        \toprule
        & \multicolumn{2}{c}{Yb--C\;\;\;\;}  & C$\equiv$C & \multicolumn{2}{c}{ C--Ca} & \\
        \hline
         State &  $L_0$ & $E_0$ &   $L_0$  &  $L_0$  &  $E_0$  & $||\mu||$ \\
          & (\AA)  &  (eV) & (\AA) & (\AA) & (eV) & (Debye) \\
         \hline
         $X~^3\Sigma$ & 2.351 &  6.5209 & 1.243 & 2.290 & 4.5179 & 0.9715 \\
         Ca $A ~\Pi$ & 2.352 &  4.6355  & 1.243 & 2.282 &  2.6336 & -- \\
         Yb $A ~\Pi$ & 2.306 &  4.1175  & 1.241 & 2.298 &  2.1145  & -- \\
         Ca $B ~^3\Sigma$ & 2.347  &  4.1918  &  1.242  &  2.295  &  2.1888 & -- \\ 
         \bottomrule
    \end{tabular}
    \caption{YbCCCa bond lengths $L_0$, bond energies $E_0$, and molecular frame permanent dipole moment ($\mu$) for the ground state, along with bond lengths in the excited states of interest.}
    \label{tab:bondlengthYbCCCa}
    
\end{table}

Similar to the $X$ ground state, close competition between singlet and triplet spin configurations is also observed in the $A$ and $B$ states. A full treatment of these excited states including spin-orbit coupling effects reveals that the small energy gap between $A^1\Pi$ and $A^3\Pi$ states induces strong intersystem crossing, causing some of the true $A\Pi$ spin-orbit sub-levels to be linear combinations of both $^1\Pi$ and $^3\Pi$ configurations~\footnote{The $m_s$=0 sectors of the singlet and triplet manifolds combine to form heavily mixed spin-orbit sub-levels of the $A~\Pi$ state. The other sub-levels are linear combinations of states in the $m_s = \pm 1$ sectors, meaning they are entirely within the triplet manifold.}. Such mixing does not occur between the $B$ state sub-levels due to their $\Sigma$ symmetry. Table \ref{tab:trueLevelsYbCCCa} gives the strength of the mixing for the $A$ states, along with the excitation energies and transition dipole moments ($\tdm$) for all the states of interest. Although the mixing reported in the Table is derived from the exact value of the small energy gap between $^1\Pi$ and $^3\Pi$ configurations, our general conclusion of ``strong spin-orbit mixing'' is robust to large relative errors in the exact value of the computed energy gap.
Further discussion of the spin-orbit splitting and mixing, including this important point, can be found in Appendix \ref{sec:SOC}.

This strong mixing between different spin configurations is a complicating factor for any laser cooling scheme that would use Ca $X\to A$ as the main transition. Electrons which are pumped into the mixed $A$ sub-levels have a roughly equal chance of decaying back into the singlet or triplet manifold of the ground state. Since these two manifolds are not exactly degenerate, they each have their own set of vibrational modes which are essentially identical to each other, but are split by $\sim 100$ GHz. This doubles the number of accessible vibrational states during an $A\Pi \rightsquigarrow X$ decay. To avoid doubling the number of repump lasers, one would need to selectively pump the $X^3\Sigma$ ground state and use the exclusively triplet $X^3\Sigma \to A^3\Pi$ transition for laser cooling. The number of additional repump lasers needed for this selective pumping of $X^3\Sigma$, on top of those necessary for the $X^3\Sigma$ vibrational states, depends heavily on the precise values of the main FCFs for the $A~\Pi \rightsquigarrow X^1\Sigma$ decay. Note, however, that these strong couplings in the $A$ states could be useful for engineering couplings between the centers.

The Ca $X \to B$ transition provides a simpler laser cooling scheme, as the $\Sigma$ symmetry of both $X$ and $B$ prevents their spin-orbit sub-levels from mixing. One could therefore drive the $X^1\Sigma \to B^1\Sigma$ or $X^3\Sigma \to B^3\Sigma$ transition with reduced risk of leaking into the other spin manifold. Additionally, since the energy gap between the Ca $B$ and $A$ states is relatively small compared to the $B-X$ gap, the radiative decay rate for $B \rightsquigarrow A$ is suppressed by a factor of $\sim 2000$ compared the the decay rate for $B \rightsquigarrow X$. 

\begin{table}[t]
\begin{center}
    \begin{tabular}{rcl@{\hskip 1mm}cc@{\hskip 1mm}c@{\hskip 1mm}c}
    \toprule
         \multicolumn{3}{c}{Transition} & \multicolumn{2}{c}{Energy} & $||\tdm||$  & $^1\Pi/^3\Pi$  \\
         & & & (eV) & (nm) & (Debye) & Admix (\%)  \\
         \hline
         $X~^1\Sigma$& $\rightarrow$ & $\mathrm{Ca} ~A~\Pi$\hspace{1mm} & 1.8843 & 658  & 4.020 & 52\%/48\% \\
         Ca $A~\Pi$\hspace{0mm} & $\rightsquigarrow$ & \hspace{-0mm}$~X~^3\Sigma$ & 1.8834 & 658 & 4.208 & 52\%/48\% \\
         $X~^3\Sigma$& $\to$ & Ca $A~^3\Pi$\hspace{1mm} & 1.8824 & 657 & 4.031 & 0\%/100\% \\
         $X~^3\Sigma$\hspace{-0mm} & $\rightarrow$ & \hspace{-0mm}$~B~^3\Sigma$ & 2.3314 & 532 & 4.747 & --  \\
         $X~^1\Sigma$\hspace{-0mm} & $\rightarrow$ & \hspace{-0mm}$~B~^1\Sigma$ & 2.3291 & 532 & 4.718 & -- \\
         $X~^1\Sigma$\hspace{-0mm} & $\rightarrow$ & \hspace{-0mm}$\mathrm{Yb} ~A~\Pi$ & 2.4034 & 516 & 4.560 & 50\%/50\% \\
         Yb $A~\Pi$\hspace{-0mm} & $\rightsquigarrow$ & \hspace{-0mm}$~X~^3\Sigma$ & 2.4025 & 516 & 4.576 & 50\%/50\% \\
         $X~^3\Sigma$ & $\to$ & Yb $A~^3\Pi$\hspace{1mm} & 2.4021 & 515 & 4.558 & 0\%/100\% \\
         $X~^3\Sigma$\hspace{-0mm} & $-$ & \hspace{-0mm}$~X~^1\Sigma$ & 0.0009 & -- & -- & --  \\
         $B~^3\Sigma$\hspace{-0mm} & $-$ & \hspace{-0mm}$~B~^1\Sigma$ & 0.0032 & -- & -- & --  \\
         \bottomrule
    \end{tabular}
\end{center}
    \caption{Computed excitation energies and transition dipole moments for the lowest-lying excitations of YbCCCa. Here $X$ denotes the ground state and $A$, $B$ are the excited states. Since the spin-orbit effects are strong in this molecule, some of the $A~\Pi$ spin-orbit sub-levels cannot be identified as purely singlet or triplet (they are left without a spin label). The other $A~\Pi$ sub-levels are purely triplet states, so they are identified with the proper label. Column 4 details the magnitude of the singlet and triplet components of the mixed $A~\Pi$ states. Detailed discussion of spin-orbit mixing is provided in Appendix \ref{sec:SOC}.}
    \label{tab:trueLevelsYbCCCa}
    
\end{table}

Note that for both molecules examined in this work, we consider only single excitations.  Excitation of one of the metal centers will shift the transitions, likely on the order of the change in spin-orbit splitting between ground and excited states. Given that this shift is considerably larger than the radiative width~\footnote{even the energy of interaction between the spin and orbital magnetic moments of the different metal-centered electrons is likely larger than the radiative width}, single excitations therefore likely ``blockade'' a second excitation, making simultaneous excitations difficult to achieve in the laboratory without the addition of even more lasers.  On the other hand, these effects likely have interesting applications on their own.

\subsection{YbCCAl}
\label{sec:ybccal_electrons}
\begin{figure*}
  \begin{minipage}[c]{0.65\textwidth}
    \begin{center}
\includegraphics[width=\textwidth]{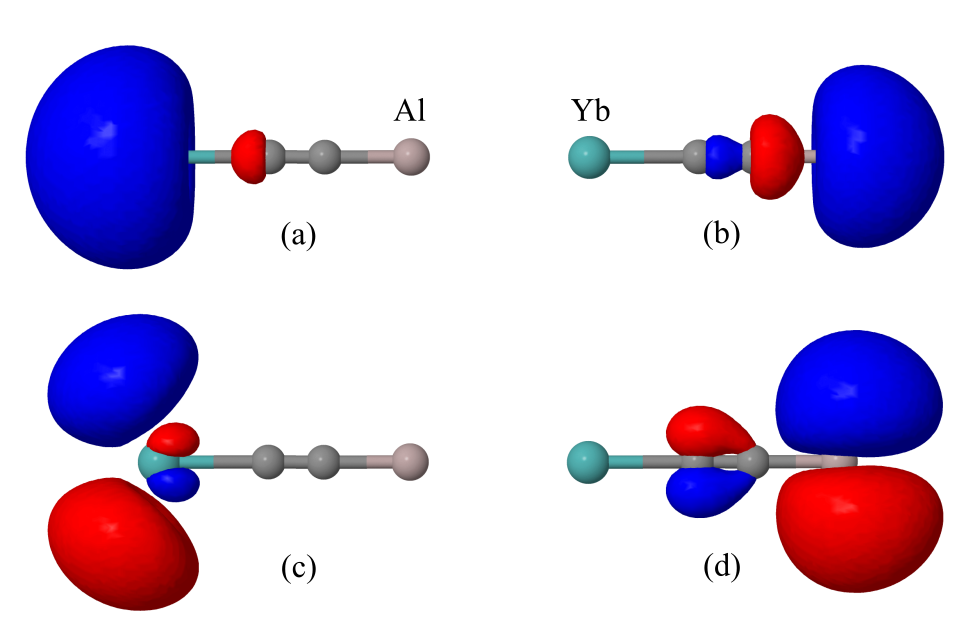}
\end{center}
  \end{minipage}\hfill
  \begin{minipage}[c]{0.30\textwidth}
    \caption{Electron and hole orbitals for the ground state and low lying excited states of YbCCAl, computed as the eigenvectors of the difference of density matrices $\rho_{\mathrm{ES}} - \rho_{\mathrm{GS}}$. \emph{Top:}  The hole orbitals, each of which is occupied in the ground state and unoccupied in the $\Pi$ excited state that is depicted directly below it. (a) shows the unpaired $6s\Sigma$ HOMO-1 on the Yb atom, while (b) shows the $3s\Sigma$ HOMO on the Al atom.  \emph{Bottom:}  The lowest-lying excited state ($A\Pi$) for the Yb $\Sigma$ electron (c), and the Al $\Sigma$ electron (d).} \label{fig:orbitalsYbCCAl}
  \end{minipage}
\end{figure*}

Given the desirable electronic transition structure and orbital hybridization of YbCCCa, but the complicated spin structure, the molecule YbCCAl has also been investigated. Since this molecule only has a doublet spin configuration in both the ground and excited states, the intersystem crossing issues in YbCCCa are avoided in YbCCAl. Similar to YbCCCa, we find that the energy of the linear geometry of YbCCAl is lower than bent and trigonal structures for both the $X^2\Sigma$ ground and $A^2\Pi$ excited states, which is supported by spectroscopy on similar molecules~\cite{Loock1997,apetrei2007gas}. Also, the excited states of interest again lie below the ionization energy. As before, the bond lengths, bond energies and permanent dipole moment are given in Table \ref{tab:bondlengthYbCCAl}, the excited state energies and transition dipole moments are given in Table \ref{tab:trueLevelsYbCCAl}, and the electron/hole orbitals for the ground state and low lying excited states are given in Figure \ref{fig:orbitalsYbCCAl}.

\begin{table}[t]
    \centering
    \begin{tabular}{ccc@{\hskip 4mm}c@{\hskip 4mm}ccc}
        \toprule
        & \multicolumn{2}{c}{Yb--C\;\;\;\;}  & C$\equiv$C & \multicolumn{2}{c}{C--Al} & \\
        \hline
         State &  $L_0$  & $E_0$ &   $L_0$ &  $L_0$  &  $E_0$ & $||\mu||$  \\
         & (\AA) & (eV) & (\AA) & (\AA) & (eV) & (Debye)\\
         \hline
         $X~^2\Sigma$ & 2.448 &  4.4757  &  1.228  & 1.935 & 5.8613 & 4.3541 \\
         Al $A ~^2\Pi$ & 2.455 &  2.0163  &  1.218  & 1.895 &  3.4019  & -- \\
         Yb $A ~^2\Pi$ & 2.401 &  2.2828  & 1.221  & 1.952 &  3.6684  & -- \\
         \bottomrule
    \end{tabular}
    \caption{YbCCAl bond lengths $L_0$, bond energies $E_0$, and molecular frame permanent dipole moment ($\mu$) for the ground state, along with bond lengths in the excited states of interest.}
    \label{tab:bondlengthYbCCAl}
    
\end{table}

\begin{table}[t]
\begin{center}
    \begin{tabular}{c@{\hskip 4mm}cc@{\hskip 4mm}c}
    \toprule
         Transition & \multicolumn{2}{c}{Energy\;\;\;\;} & $||\tdm||$  \\
         & (eV) & (nm) & (Debye) \\ \hline
         $X~^2\Sigma \rightarrow \mathrm{Al} ~A~^2\Pi$ & 2.4594 & 504  & 0.0735 \\
         $X~^2\Sigma \rightarrow \mathrm{Yb} ~A~^2\Pi$ & 2.1929 & 565 & 6.345 \\
         \bottomrule
    \end{tabular}
\end{center}
    \caption{Computed excitation energies and transition dipole moments for the lowest-lying excitations of YbCCAl. Here $X$ denotes the ground state and $A$ denotes an excited state.}
    \label{tab:trueLevelsYbCCAl}
    
\end{table}

We find that YbCCAl is also open-shell and has the desired bonding pattern, which causes the ground state to have a single, unpaired, non-bonding $6s\Sigma$ electron on the Yb as the HOMO and two non-bonding $3s\Sigma$ electrons on the Al as the dominant feature of the HOMO-1 (see Fig. \ref{fig:orbitalsYbCCAl}a-b). The $X\to A$ excited state on the Yb atom (Fig. \ref{fig:orbitalsYbCCAl}c) is highly similar to the $X\to A$ excited state in YbCCCa (Fig. \ref{fig:orbitalsYbCCCa}c), as expected.  

However, the $X\to A$ excitation on the Al atom (Fig. \ref{fig:orbitalsYbCCAl}d) has noticeably worse features than the corresponding Ca excitation in YbCCCa (Fig. \ref{fig:orbitalsYbCCCa}d). The lack of $s-p$ hybridization of the Al orbitals leaves a significant portion of the electron density in the Al $A$ state within the C-Al bonding region. This leads to a relatively large change in the C-Al bond length (Table \ref{tab:bondlengthYbCCAl}) when compared to the corresponding C-Ca bond length in the Ca $A$ state of YbCCCa (Table \ref{tab:bondlengthYbCCCa}). Further, the Al $A$ state is not completely localized on the Al atom, as there is some additional density in the $\pi$-bonding region of the C$\equiv$C bond. This induces a shortening of the C$\equiv$C bond, similar to what was observed experimentally in AlCCH~\cite{apetrei2007gas}, which we do not observe in the excited states of YbCCCa. These features of the Al $A$ state are not optimal for photon cycling, as geometry changes tend to reduce the diagonality of the FC matrix.

As a final note, we point out that for both YbCCCa and YbCCAl our calculations reveal that there are no $\Delta$ excited states on the metal atoms which are intermediate in energy between the $\Sigma$ and $\Pi$ states we have reported. This result is expected for the Al atom, but the level structures of atomic Yb$^+$ and Ca$^+$ would suggest intermediate $\Delta$ states. This ``reordering" of the $\Pi$ and $\Delta$ levels has been understood for alkaline earth monohalides using a ligand field model \cite{rice1985electronic,allouche1993ligand}, as well as for CaOH using electronic structure techniques \cite{taylor2005electronic}. It has also been observed experimentally~\cite{Verges1993,Jarman1992}. We expect that the qualitative nature of these results hold for our slightly more complicated YbCCM species. However, there may still be perturbative correlated effects between the low-lying $\Pi$ states and the displaced $\Delta$ states which have physical implications that are not captured in this study \cite{Melville2001}. 

\section{Vibrational Structure}

FC matrix elements have been computed for both YbCCCa and YbCCAl, using methods described in Section \ref{sec:compDetails}. There we also discuss that the typical level of error in these calculations is no less than a few percent, and thus unequivocal assessment of the cycling properties of these molecules always requires experimental measurements. Nonetheless, the numbers presented in this section serve as a useful guide for future experimental and theoretical investigation of these exotic hypermetallic small molecules.

Due to their linear geometry, YbCCCa and YbCCAl both have five vibrational modes, two of which are doubly-degenerate, for a total of 7 modes. The assignments and energies of these modes are given for the triplet ground states of both molecules in Table \ref{tab:frequencies_both}.   $\nu_1$, $\nu_2$, and $\nu_3$ label the number of vibrational quanta populating the C$-M$, Yb$-$C, and C$\equiv$C stretching modes of the molecule YbCCM, respectively (where M is either Ca or Al). These modes are non-degenerate and have the same $\sigma$ symmetry as the vibronic ground state. In reality the physical modes are superpositions of these possible modes, though these simple descriptions are reasonably accurate due to the mass differences between the constituent atoms~\cite{Oberlander1995,Kozyryev2017PolyEDM}. $\nu_4$ and $\nu_5$ label the population of the doubly-degenerate ``anti-symmetric" and ``symmetric" bending modes of the C atoms about the symmetry axis, respectively~\footnote{ Since the YbCCM molecules have no inversion symmetry, the anti-symmetric and symmetric nomenclature to distinguish the bending modes technically does not make sense. However, these names are well-defined for 4 atom molecules with inversion symmetry, such as HCCH, and there is a very clear correspondence between the normal modes of YbCCM and those of HCCH, so we use this nomenclature anyways for the sake of clarity.}.  Linear combinations of these $\pi$-symmetric degenerate pairs can be formed which correspond to states with definite angular momentum $\ell_k = \nu_k,\nu_k-2,\nu_k-4,..., 1/0$ about the symmetry axis~\cite{Herzberg1967,Kozyryev2017PolyEDM}. This quantum number $\ell$ characterizes the different sub-levels which can occur for the degenerate modes.

\begin{table}[t]
\centering
\begin{tabular}{cc@{\hskip 4mm}r@{\hskip 4mm}r}
\toprule
Mode & Assignment & \multicolumn{2}{c}{Frequency (cm$^{-1}$)} \\
 & & YbCCCa & YbCCAl \\
\hline
C$-$M Stretch & $\nu_1(\sigma)$ &  484.92\hspace{1mm} & 612.78\hspace{1mm}  \\
Yb$-$C Stretch & $\nu_2(\sigma)$ & 185.57\hspace{1mm} & 216.77\hspace{1mm} \\
C$\equiv$C Stretch & $\nu_3(\sigma)$ & 2006.15\hspace{1mm} & 2047.00\hspace{1mm} \\
Asymm. Bend & $\nu_4(\pi)$ & 43.15\hspace{1mm} & 58.34\hspace{1mm}  \\
Symm. Bend & $\nu_5(\pi)$  & 140.89\hspace{1mm} & 197.60\hspace{1mm}  \\
\bottomrule
\end{tabular}
\caption{Vibration frequencies for each of the vibrational modes of the triplet ground state of YbCCM, where M is either Ca or Al.  $\sigma$ and $\pi$ represent the symmetry of the vibration, and whether it is non-degenerate or doubly-degenerate, respectively.}
\label{tab:frequencies_both}

\end{table}

We will denote the vibrational wavefunction of the ground electronic state $\chi_0(S)$, where $S=\{\nu_1\nu_2\nu_3\nu_4\nu_5\}$ is some vibrational state.  Similarly, $\chi_1(0)$ will denote the excited electronic state.  We only consider decays from the ground vibrational state in the excited electronic state since molecules will be excited to this state selectively during the laser cooling process, though other excitations are relevant for laser cooling schemes since repumping is invariably required.  We seek to compute the FCFs $|\langle \chi_0(S) | \chi_1(0) \rangle|^2$ for the relevant $\ket{\chi_1}$ states discussed earlier in order to understand if photon cycling on the two metal atoms is possible with only a small number of repumping lasers~\footnote{Note that the original literature calls \unexpanded{$\langle \chi_0(S) | \chi_1(0) \rangle$} the Franck-Condon Factor, but that definition is less useful and increasingly uncommon}. 

In general, the restrictions on $S$ are simply that $\ket{\chi_0(S)}$ must have a total symmetry of $\sigma$, because the $\ket{\chi_1(0)}$ state is $\sigma$-symmetric and therefore any other spontaneous decays $\ket{\chi_1(0)} \rightsquigarrow \ket{\chi_0(S)}$ are forbidden. More specifically, this means that the stretching modes $\nu_{1-3}$ can be arbitrarily populated, but the bending modes $\nu_{4-5}$ are subject to selection rules during radiative decay of $\ket{\chi_1(0)}$~\cite{Herzberg1967},
\begin{gather}
\Delta \ell_k = 0, ~~k \in \{4,5\} \nonumber\\
\sum_{k} \Delta \nu_k = 0,\pm2,\pm4,\pm6,...
\end{gather}
Finally, for any single excited state $\ket{\chi_1(0)}$ under consideration in this work we have the useful property,
\begin{equation}
\sum_S|\langle \chi_0(S) | \chi_1(0) \rangle|^2 = 1, \label{eqn:FCF}
\end{equation}
which provides a normalized scale with which to assess the branching ratios.

\subsection{YbCCCa}
\label{sec:ybccca_fcfs}

\begin{table}[t]
\begin{center}
\begin{tabular}{c@{\hskip 4mm}l@{\hskip 4mm}l}
\toprule
 $X^3\Sigma \rightarrow \mathrm{Ca}~A^3\Pi$ & FCF  & Sum \\
\hline
$\chi_0(\{ 00000 \}) \rightarrow \chi_1(0)$ & 0.99 & 0.99 \\
$\chi_0(\{ 01000 \}) \rightarrow \chi_1(0)$ & 0.005 & 0.994 \\
$\chi_0(\{ 10000 \}) \rightarrow \chi_1(0)$ & 0.003 & 0.997 \\
$\chi_0(\{ 0002^0 0 \}) \rightarrow \chi_1(0)$ & 0.002 & 0.999 \\
$\chi_0(\{ 00011 \}) \rightarrow \chi_1(0)$ & 0.0004 & 0.9994 \\
$\chi_0(\{ 20000 \}) \rightarrow \chi_1(0)$ & 0.0003 & 0.9997 \\
$\chi_0(\{ 11000 \}) \rightarrow \chi_1(0)$ & 0.0002 & 0.9999 \\
\hline
Sum: & 0.9999 \\
\midrule[1.5pt]
$X^3\Sigma \rightarrow \mathrm{Yb}~A^3\Pi$ & FCF & Sum\\
\hline
$\chi_0(\{ 00000 \}) \rightarrow \chi_1(0)$ & 0.75 & 0.75 \\
$\chi_0(\{ 01000 \}) \rightarrow \chi_1(0)$ & 0.16 & 0.91 \\
$\chi_0(\{ 10000 \}) \rightarrow \chi_1(0)$ & 0.06 & 0.97 \\
$\chi_0(\{ 02000 \}) \rightarrow \chi_1(0)$ & 0.01 & 0.98 \\
$\chi_0(\{ 11000 \}) \rightarrow \chi_1(0)$ & 0.01 & 0.99 \\
$\chi_0(\{ 20000 \}) \rightarrow \chi_1(0)$ & 0.004 & 0.995 \\
$\chi_0(\{ 03000 \}) \rightarrow \chi_1(0)$ & 0.0008 & 0.996 \\
$\chi_0(\{ 00011 \}) \rightarrow \chi_1(0)$ & 0.0003 & 0.996 \\
\hline
Sum: & 0.996  \\
\midrule[1.5pt]
$X~^3\Sigma \rightarrow \mathrm{Ca}~B^3\Sigma$ & FCF & Sum\\
\hline
$\chi_0(\{ 00000 \}) \rightarrow \chi_1(0)$ & 0.995 & 0.995 \\
$\chi_0(\{ 10000 \}) \rightarrow \chi_1(0)$ & 0.003 & 0.998 \\
$\chi_0(\{ 00002^0 \}) \rightarrow \chi_1(0)$ & 0.001 & 0.999 \\
$\chi_0(\{ 0002^0 0 \}) \rightarrow \chi_1(0)$ & 0.0004 & 0.9994 \\
$\chi_0(\{ 20000 \}) \rightarrow \chi_1(0)$ & 0.0003 & 0.9997 \\
$\chi_0(\{ 11000 \}) \rightarrow \chi_1(0)$ & 0.0002 & 0.9999 \\
\hline
Sum: & 0.9999  \\
\bottomrule
\end{tabular}
\caption{Franck-Condon factors for the metal-centered electronic transitions of YbCCCa.
Multiply populated vibrational modes $\nu$ which are degenerate bending modes
are additionally labeled by their symmetry-projected
angular momentum quantum number $\nu^{\ell}$. Effects of systematic errors in the
calculations are not included in these numbers (see Sections \ref{sec:ybccca_fcfs} 
\& \ref{sec:compDetails} for estimates of these effects).}
\label{tab:FCF_YbCCCa}
\end{center}
\end{table}

The computed FCFs for the $X^3\Sigma \to \mathrm{Ca}~A^3\Pi$, $X^3\Sigma \to \mathrm{Yb}~A^3\Pi$, and $X^3\Sigma \to \mathrm{Ca}~B^3\Sigma$ transitions are shown in Table \ref{tab:FCF_YbCCCa}. The FC matrix for the Ca $X\to A$ transition was calculated to be highly diagonal, with a 0-0 FCF of 0.99 and only 3 other transitions with an FCF greater than $4 \cdot 10^{-4}$. However, quantitative estimates of the systematic error in these calculations suggest a reduced level of diagonality. Accounting for our ``worst case" error estimates for this transition, the 0-0 FCF is reduced to 0.9, but only 3 additional states are required to reach a total efficiency of 0.997, which is comparable with the ``diagonal" results in Table \ref{tab:FCF_YbCCCa}. However, 8 total states (1 main transition + 7 repump states) are required for an efficiency of 0.999. which is significantly worse than the results in Table \ref{tab:FCF_YbCCCa}. The details of these error estimates are discussed in Section \ref{sec:compDetails}.

Without considering the estimates of systematic error or the complications that arise due to intersystem crossing, the results in Table \ref{tab:FCF_YbCCCa} suggest that the Ca atom can scatter thousands of photons with only 3-4 lasers (1 main transition and 2-3 repumps) and tens of thousands of photons with 7 lasers. When considering the worst case systematic error, these numbers are increased to 4-8 lasers just for the ability to scatter thousands of photons. Further, the consideration of intersystem crossing effects requires the addition of even more lasers in order to have a ``closed" laser cooling cycle that accounts for the non-degenerate vibrational manifolds of the $X^1\Sigma$ and $X^3\Sigma$ states (as discussed in Section \ref{sec:ybccca_electrons}).

The Ca $X\to B$ displays significantly nicer properties. The FC matrix was computed to be even more diagonal than the $X\to A$ transition, with a 0-0 FCF of 0.995 and only 2 other transitions with an FCF greater than $4 \cdot 10^{-4}$. Additionally, the effects of the ``worst case" systematic error estimates for this transition are smaller: the 0-0 FCF is reduced to 0.993, 4 total states are required for an efficiency of 0.9991, and 7 total states are required for an efficiency of 0.9999.  Further, intersystem crossing effects are suppressed. This allows the Ca center to scatter thousands of photons with only 3-4 lasers, and tens of thousands of photons with 6-7 lasers. 

The Yb-centered transition is less diagonal, showing more expansive branching than YbOH~\cite{Kozyryev2017PolyEDM}. The main 0-0 FCF is only 0.75 and there are 5 other FCFs with values larger than $10^{-3}$. This limits the scattering efficiency of the Yb atom to $\sim 500$ photons with a reasonable number of repump lasers, without even considering systematic errors or the additional lasers that are necessary due to intersystem crossing. This decreased efficiency of the Yb atom compared to YbOH is likely due to the fact that the Yb--C bond in YbCCCa is significantly longer and ``floppier'' than the Yb--O bond in YbOH. This allows for a more significant off-diagonal vibrational decay channel through the Yb--C stretch, as we can see in Table \ref{tab:FCF_YbCCCa}. Note that for each metal center M, the two most dominant off-diagonal decays for the $A$ states are the M--C stretch and the C--M$'$ stretch. The former is not surprising, but the latter may seem unusual since the metal centers are rather far apart. However, the descriptions of the mode assignments shown in Table \ref{tab:frequencies_both} are only an approximation, and the true physical normal modes are admixtures of the idealized stretching modes described in the first column.

\subsection{YbCCAl}

The FCFs for the $X^2\Sigma \rightarrow \mathrm{Al}~A^2\Pi$ and $X^2\Sigma \rightarrow \mathrm{Yb}~A^2\Pi$ transitions are shown in Table \ref{tab:FCF_YbCCAl}. The Yb $X\to A$ excitation shows similar branching ratios to the Yb-centered excitation in YbCCCa, although the populated modes differ slightly.  Despite the same branching ratios, the YbCCAl excitation will have a higher optical efficiency in practice because it avoids the intersystem crossing of the YbCCCa $A$ state. On the other hand, the Al-centered excitation is significantly less diagonal than the Ca $X\to A$ transition in YbCCCa. The large number of significant FCFs cause its optical efficiency to be too low for successful laser cooling. This result is expected based on the non-ideal electronic density in the Al $A^2\Pi$ excited state (Fig. \ref{fig:orbitalsYbCCAl}d) and its relatively large impact on the geometry of the molecule, as discussed in Section \ref{sec:ybccal_electrons}. 

\begin{table}[t]
\begin{center}
\begin{tabular}{c@{\hskip 4mm}l@{\hskip 4mm}l}
\toprule
 $X~^2\Sigma \rightarrow \mathrm{Al}~A~^2\Pi$ & FCF  & Sum \\
\hline
$\chi_0(\{ 00000 \}) \rightarrow \chi_1(0)$ & 0.74 & 0.74 \\
$\chi_0(\{ 10000 \}) \rightarrow \chi_1(0)$ & 0.14 & 0.88 \\
$\chi_0(\{ 01000 \}) \rightarrow \chi_1(0)$ & 0.06 & 0.95 \\
$\chi_0(\{ 11000 \}) \rightarrow \chi_1(0)$ & 0.01 & 0.96 \\
$\chi_0(\{ 00002^0 \}) \rightarrow \chi_1(0)$ & 0.01 & 0.97 \\
$\chi_0(\{ 20000 \}) \rightarrow \chi_1(0)$ & 0.01 & 0.98 \\
$\chi_0(\{ 00011 \}) \rightarrow \chi_1(0)$ & 0.009 & 0.984 \\
$\chi_0(\{ 00100 \}) \rightarrow \chi_1(0)$ & 0.005 & 0.989 \\
$\chi_0(\{ 02000 \}) \rightarrow \chi_1(0)$ & 0.002 & 0.991 \\
\hline
Sum: & 0.991 \\
\midrule[1.5pt]
$X~^2\Sigma \rightarrow \mathrm{Yb}~A~^2\Pi$ & FCF & Sum\\
\hline
$\chi_0(\{ 00000 \}) \rightarrow \chi_1(0)$ & 0.74 & 0.74 \\
$\chi_0(\{ 01000 \}) \rightarrow \chi_1(0)$ & 0.18 & 0.91 \\
$\chi_0(\{ 10000 \}) \rightarrow \chi_1(0)$ & 0.05 & 0.96 \\
$\chi_0(\{ 02000 \}) \rightarrow \chi_1(0)$ & 0.02 & 0.98 \\
$\chi_0(\{ 11000 \}) \rightarrow \chi_1(0)$ & 0.009 & 0.988 \\
$\chi_0(\{ 00100 \}) \rightarrow \chi_1(0)$ & 0.005 & 0.993 \\
$\chi_0(\{ 20000 \}) \rightarrow \chi_1(0)$ & 0.002 & 0.995 \\
$\chi_0(\{ 10100 \}) \rightarrow \chi_1(0)$ & 0.001 & 0.996 \\
\hline
Sum: & 0.996  \\
\bottomrule
\end{tabular}
\caption{Franck-Condon factors for the metal-centered electronic transitions of YbCCAl.
Multiply populated vibrational modes $\nu$ which are degenerate bending modes
are additionally labeled by their symmetry-projected
angular momentum quantum number $\nu^{\ell}$.}
\label{tab:FCF_YbCCAl}
\end{center}
\end{table}

\section{Computational Details}
\label{sec:compDetails}

The molecules YbCCCa and YbCCAl were predominantly investigated with the complete active space self-consistent field (CASSCF)~\cite{werner1985second,knowles1985efficient} and multireference configuration interaction (MRCI)~\cite{werner1988efficient,knowles1988efficient} methods from the MOLPRO quantum chemistry package~\cite{MOLPRO_brief}. Since we were interested in the nature of the low-lying excited states, all calculations involving excited states were performed using state averaging in CASSCF. These methods were chosen in order to address the electronic levels (including spin-orbit effects) as accurately as possible, despite the inherent difficulty of using them to compute more challenging quantities such as the FC matrix elements (due to very high computational cost). 

Despite their typical levels of accuracy, using active space-based methods with heavy atoms such as Yb poses a challenge. As an example, consider the ideal chemical active space for YbCCCa: it should likely include all doubly occupied valence $f$ orbitals on the Yb, 4 doubly occupied bonding orbitals, 2 singly occupied valence $s\Sigma$ orbitals, 4 virtual $p\Pi$ orbitals, and valence $d$ orbitals on both the Yb and Ca. This active space (24 electrons, 28 orbitals) is far too large for MRCI, and even if the virtual $d$ orbitals are removed, the (24e, 18o) reduced active space is still too large for MRCI. 

To test whether this challenge would prohibit us from using the MRCI methodology (in favor of a more approximate, less computationally expensive method), we examined the importance of including the occupied $f$ orbitals and virtual $d$ orbitals in the active space by running MRCI on further reduced ``test'' active spaces that were constructed by exploiting orbital symmetries. Specifically, some essential components of the test active spaces always remained the same: the 4 bonding orbitals, 2 singly occupied valence $s\Sigma$ orbitals (a-b in Figs. \ref{fig:orbitalsYbCCCa} \& \ref{fig:orbitalsYbCCAl}), the 4 virtual $p\Pi$ orbitals (c-d in Figs. \ref{fig:orbitalsYbCCCa} \& \ref{fig:orbitalsYbCCAl}), and the Ca $3d\Sigma$ orbital (Fig. \ref{fig:orbitalsYbCCCa}e). However, added on top of these were permutations of the occupied $f$ orbitals in 0, 1, or 2 symmetry sectors and the valence $d$ orbitals in 0, 1, or 2 additional symmetry sectors (not necessarily the same as the $f$ sectors). This allowed us to perform well-defined MRCI tests (ie. the exact orbitals in the active space were unambiguously known) on active spaces of more reasonable sizes such as (14e, 14o) - (16e, 16o). For example, a prototypical (14e, 14o) test space included 2 occupied $f$ orbitals and one virtual $d$ orbital on each metal, along with the 4 bonding orbitals, 2 singly occupied valence $s\Sigma$ orbitals, and the 4 virtual $p\Pi$ orbitals.

Across many different permutations of such active spaces, the MRCI tests revealed that configurations containing occupied valence $d$ orbitals maximally contributed $\sim 0.02\%$ to the multi-determinant ground state and excited $A$ states of interest. The $B$ state in YbCCCa was completely dominated by the occupation of the $\Sigma$-symmetric Ca $3d$ orbital in Fig. \ref{fig:orbitalsYbCCCa}e, and had equally small contributions from the $3d$ orbitals in other symmetry groups. This was likely due to the relatively large energy gap between the $3d\Sigma$ orbital and the $3d \Pi,\Delta$ orbitals, arising from the significant $s-d$ and $p_z-d$ hybridization seen in Fig. \ref{fig:orbitalsYbCCCa}e (and the lack of any hybridization in the other $d$ orbitals).
A completely negligible contribution arose from all configurations in which an electron vacated the doubly occupied $f$ or bonding orbitals. This suggests that, at least in this specific case, molecular Yb has significantly simpler electronic structure than atomic Yb$^+$~\cite{porsev2012correlation,feldker2018spectroscopy}, as mentioned at the end of Section \ref{sec:ybccal_electrons}. 

The excited states were examined in thse tests by including 12 baseline states in the state averaging procedure: 2 $^1\Sigma$, 2 $^3\Sigma$, 4 $^1\Pi$, and 4 $^3\Pi$. Additional states were added to this average based on the symmetry of the $f$ and $d$ orbitals included in a given active space permutation. These tests allowed us to search for all the low-lying excited states and revealed no additional allowed states in the energy range of the $A$ and $B$ states of interest. Most importantly, the results allowed us to define much more tractable (yet still physically realistic) active spaces for our MRCI studies of YbCCCa and YbCCAl. 

\subsection{YbCCCa}

\begin{figure}
    \centering
    \includegraphics[width=\columnwidth]{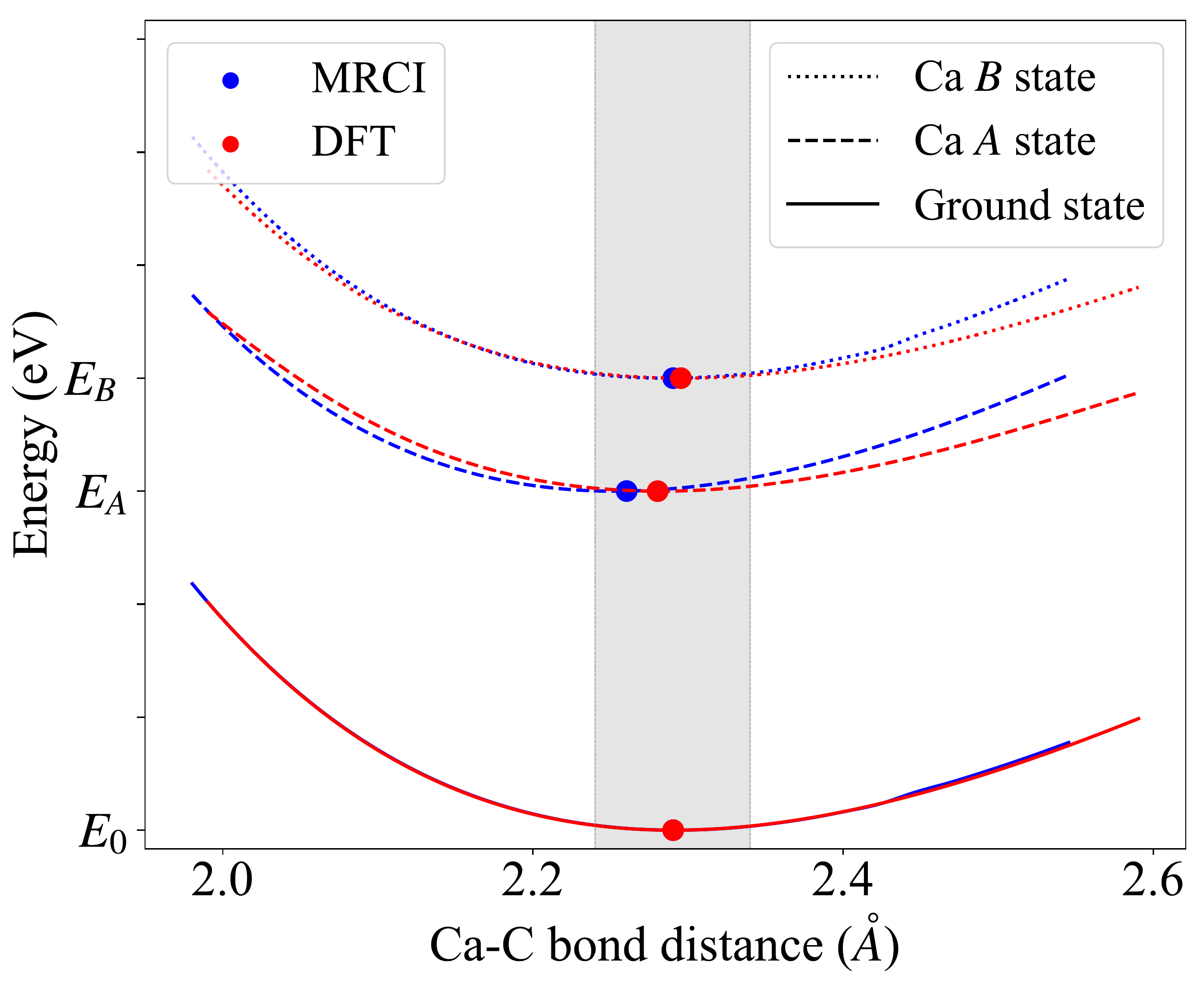}
    \caption{Potential energy surfaces obtained with DFT and MRCI along the Ca-C bond coordinate in YbCCCa. The solid lines are the ground state, dashed lines are the Ca $A$ state, and dotted lines are the Ca $B$ state. Red curves are from DFT and blue curves are from MRCI. The dots show the minima of each curve. The curves have been shifted by scalar values in order to make them easier to compare; hence the energy axis does not have numerical tick labels because the gaps between the curves are not to scale. However, for a sense of scale of each curve individually, the tick marks are placed at intervals of 0.3 eV.  }
    \label{fig:dftVmrci}
\end{figure}

All electronic orbitals, density matrices, transition dipole moments, excitation energies, and bond energies were obtained using MRCI. The full ANO-RCC basis was used for Ca and C~\cite{roos2004relativistic,roos2004main} while a contracted ANO basis~\cite{cao2001valence,cao2002segmented} was used in combination with a 28 electron small core pseudopotential~\cite{dolg1989energy} for the treatment of scalar relativistic effects in Yb. An active space of (2e, 7o) was used, including the two singly occupied valence $s\Sigma$ electrons (Fig. \ref{fig:orbitalsYbCCCa}a-b), 4 virtual $p\Pi$ orbitals (Fig. \ref{fig:orbitalsYbCCCa}c-d), and the one relevant Ca $3d\Sigma$ virtual orbital shown in Fig. \ref{fig:orbitalsYbCCCa}e. Given the active space, we used 12 states in the state averaging procedure: 2 $^1\Sigma$, 2 $^3\Sigma$, 4 $^1\Pi$, and 4 $^3\Pi$. The spin-orbit coupling analysis was done in MOLPRO via the state interaction formalism \cite{malmqvist2002restricted,sjovoll1997determinantal,berning2000spin,yabushita1999spin} with all 12 of the MRCI wavefunctions from the state averaging.

Optimized geometries, normal modes, and vibrational frequencies for the ground and excited states were obtained at the level of unrestricted Kohn-Sham (UKS) DFT using the B3LYP functional in the Q-Chem quantum chemistry package~\cite{qchem}. Excited states were obtained using TDDFT. The $X^3\Sigma$ ground state was easier to reliably isolate than the $X^1\Sigma$ ground state in the UKS procedure, so we report FC matrix elements for the triplet manifold. Given that the splitting between the singlet and triplet manifolds is on the order of $\sim 100$ GHz for the $X$, $A$, and $B$ states, we do not expect a meaningful difference in the FCFs between the singlet and triplet sub-levels of these states, up to the level of accuracy that can be expected from these calculations. This approach was chosen instead of continuing to use MRCI due to the numerical difficulty and computational cost associated with computing vibrational frequencies with CASSCF+MRCI. 

We justify the validity of this use of DFT in Figure \ref{fig:dftVmrci} by examining the DFT and MRCI energy landscapes along the relevant C-Ca bond coordinate. We use the MRCI curves to estimate the level of systematic error in the DFT frequencies and geometry changes for each state, and then assess the effects of these errors on the final FCF values. The curves in Figure \ref{fig:dftVmrci} have been shifted by scalars on both the x- and y-axes in order to make the comparison of the curves easier. Shifting in the y coordinate is necessary because DFT and MRCI do not predict the same excitation energies. Shifting in the x coordinate is required because DFT and MRCI do not predict the exact same Ca-C bond length in the ground state; however this alone has no effect on the values of the FCFs because they depend on \textit{changes} in geometry and vibrational frequencies between ground and excited states (and these quantities are preserved by scalar shifts).

The ground state curves are virtually identical in the window of importance (highlighted in gray). Their second derivatives (which are directly related to vibrational frequencies) at the equilibrium geometry differ by less than $0.5\%$ between MRCI and DFT. In the $A$ state, the $X \to A$ bond length change from the MRCI curve is $0.02$ \AA~greater than is predicted by DFT and the second derivative from MRCI is $\sim 15\%$ larger than DFT. The effects of ``worst case" systematic errors, in which we assume all elements of the DFT Hessian are underestimated by $\sim 15\%$ (along with the 0.02 \AA~Ca-C geometry change error), on the final FCF values for the Ca $X \to A$ transition are discussed in Section \ref{sec:ybccca_fcfs}. In the $B$ state, the $X \to B$ bond length change from the MRCI curve is only $\sim 0.005$ \AA~greater than is predicted by DFT, and the second derivative from MRCI is $\sim 8\%$ larger than DFT. Similarly, the effects of the ``worst case" of these errors on the FCFs are discussed in Section \ref{sec:ybccca_fcfs}.

\subsection{YbCCAl}

All electronic orbitals, density matrices, transition dipole moments, excitation energies, and bond energies were obtained with MRCI, while the optimized geometries, normal modes, and their frequencies were obtained with CASSCF (all in MOLPRO).
The def2-TZVPP basis was used for all atoms~\cite{weigend1998ri,weigend2005balanced,gulde2012error}, which included the same 28 electron small core pseudopotential as above for scalar relativistic effects. Given that the Al-centered excitation contains some C$\equiv$C $\pi$-bonding density (see Fig. \ref{fig:orbitalsYbCCAl}d), the C$\equiv$C $\pi$ bonding orbitals were included in the active space for these calculations, along with the valence $s\Sigma$ orbitals (Fig. \ref{fig:orbitalsYbCCAl}a-b) and the 4 virtual $p\Pi$ orbitals (Fig. \ref{fig:orbitalsYbCCAl}c-d) for a (7e,8o) active space. 

For this molecule, we included 5 states in the state averaging procedure: 1 $^2\Sigma$ and 4 $^2\Pi$. For the normal mode frequency calculations, we noticed less agreement between DFT and MRCI than we saw in Fig. \ref{fig:dftVmrci} for YbCCCa. This may have been due to the larger magnitude of geometry changes in the YbCCAl excited states (Section \ref{sec:ybccal_electrons}). Thus, we instead chose to use CASSCF to compute the normal modes and their frequencies.

For both molecules, the FCFs were computed from the ab initio molecular data using the ezSpectrum software~\cite{ezspectrum}. Multiple approximations enter into these calculations. Firstly, all the ab initio molecular data was obtained within the Born-Oppenheimer approximation, so certain effects such as Renner-Teller are not accounted for. Additionally, the FCFs are computed by assuming that the potential energy surfaces in the immediate vicinity of the equilibrium geometries for both the ground and excited state can be approximated by a harmonic potential. Based on Fig. \ref{fig:dftVmrci}, this does not appear to be a strong assumption in our case. Finally, the FCFs were computed analytically, including hot bands and Duchinsky rotations~\cite{kupka1986multidimensional,berger1998calculation,duschinsky1937importance}, so no assumptions were made about the normal modes of the ground and excited state being parallel. 

As a concluding remark to this section, it is worth mentioning explicitly that no quantum chemical calculations for molecules this heavy have sufficient accuracy to serve as a replacement for spectroscopic measurements.  The first experimental step toward using such molecules is measurement of the energy levels and branching ratios via broadband optical spectroscopy.

\section{Discussion}

These results suggest that both YbCCCa and YbCCAl do indeed show many of the desired properties for polar molecules with multiple cycling centers. Both the ground and excited states are bound and linear, the valence $s$ electrons of interest for laser cooling are localized on their respective metal centers, and the Franck-Condon factors are reasonably diagonal (especially for the Ca center). However, YbCCAl reveals that metals without an alkaline earth-like valence structure may not have strong enough $s-p$ hybridization of the excited state orbitals to remove the electron density from the bonding region, though this is worth investigating further with other species, particularly other group IIIA elements. It seems likely that this generic feature will often cause significant geometrical changes in the excited state of polyatomic molecules, which reduces the diagonal nature of the FC matrix and allows branching into a significant number of vibrational modes. These effects have also been discussed elsewhere in the context of molecules with a single metal center~\cite{Ellis2001}. 

In YbCCCa, both metal atoms have alkaline earth-like valence structures and the $s-p$ hybridization of the excited state orbitals significantly improves FCFs for the Ca center over the Al center. The two singly-bonded alkaline earth-like atoms create a diradical with two singly occupied $s$ orbitals, one on each metal. Since we seek a molecule which has two metal centers that are as independent as possible, this electronic structure creates a feature of critical importance: the more independent the two open-shell electrons are, the smaller the energy gap between singlet and triplet configurations of the molecule in both ground and excited states. This quasi-degeneracy of singlet and triplet states gives rise to complexities in cycling with the $X \to A$ transition. The singlet-triplet gap will likely never be small enough that it is not resolved by a laser~\footnote{Due to the relative magnitude of various Breit-Pauli terms such as the spin-spin interaction compared to the spectral width of modern lasers.}, but also not large enough that strong spin-orbit mixing of the $A^1\Pi$ and $A^3\Pi$ excited states can be avoided (even for light molecules, as is seen for Ca in Table \ref{tab:trueLevelsYbCCCa}). This could effectively double the number of vibronic states with significant FCFs, though there may be routes to avoid leakage between the singlet and triplet manifolds as discussed in Section \ref{sec:ybccca_electrons}.  

Regardless, the Ca $B$ state avoids the challenges caused by spin-orbit coupling because both the $X$ and $B$ states are $\Sigma$-symmetric. This allows the highly diagonal, spin-pure $X^1\Sigma\to B^1\Sigma$ or $X^3\Sigma\to B^3\Sigma$ transitions on the Ca center to be considered as potentially feasible laser cooling transitions. These transitions may only require $\sim 4-5$ repump lasers in order to cycle tens of thousands of photons, and the Ca $B \rightsquigarrow A$ decay has a radiative rate which is suppressed by a factor of $\sim 2000$ compared to the desired Ca $B \rightsquigarrow X$ decay, due to the difference in their respective energies.  No such $B$ state was investigated in detail on the Yb center due to the lack of existing experimental data on $B$ states in Yb-containing molecules. Additionally, configurations with holes in Yb $f$ orbitals lie energetically below any potential Yb $B$ state, which makes an accurate study of its properties significantly more challenging. 

Our results also suggest a compelling alternative approach for precision measurements utilizing hypermetallic MCCM$'$ molecules. For metals that do not make cycling centers in molecules, such as Th, Ta, U, \textit{etc}., such a molecular scheme should make it possible to cycle photons, apply optical forces, and potentially implement laser cooling to perform precision measurement on these species, while maintaining the ability to realize full polarization and internal co-magnetometer states for robust systematic error rejection~\cite{Kozyryev2017PolyEDM}.  Utilizing the diagonal transitions of a Ca (or analogous) metal center in a molecule containing Th or Ta may offer \textit{extremely favorable} coherence times compared to molecules such as ThO~\cite{Wentink1972,Kokkin2014,ACME2018} or TaN~\cite{Bouchard2016}, or polyatomic analogues such as ThOH$^+$~\cite{Flambaum2018Schiff} or TaCH~\cite{Kingston2001}.

The additional benefits of this molecular design may be numerous. First, the cycling center could be used for enhanced state detection.  By using state-dependent optical pumping (or coherent transfer) of spin states to internal states, for example excited vibrational states, the cycling center can be used for efficient detection of the initial spin state.  Second, the cycling center offers an additional co-magnetometer that can be used to diagnose stray fields and other systematic effects. Third, the requirements are more relaxed for a molecule in which the non-cycling center is the focus of the precision measurement, compared to the dual-cycling molecules examined in this work. A slight perturbation to the FCFs of an optical cycling precision measurement atom (such as Yb) can destroy experimental efficiency, but a slight perturbation to a CP-violation sensitivity parameter of a non-cycling measurement atom (such as Th) will still result in a promising molecule. Such a molecule would also avoid the ``excitation blockade'' discussed earlier since simultaneous excitation would be undesirable in the first place.  Lastly, the polyatomic structure allows us to use diamagnetic species with sensitivity to nuclear CP violation and good robustness against magnetic effects, such as the $^1\Sigma$ states of divalent Th or Ra~\cite{Kozyryev2017PolyEDM,Flambaum2018Schiff} or monovalent Tl~\cite{Hunter2012,Kozyryev2017PolyEDM} while still maintaining strong systematic error rejection and providing optical readout schemes. All of these areas are worth considering in further theoretical studies.

\section{Conclusion}

In summary, we have explored the vibronic structure of two prototypical hypermetallic small molecules for precision measurement experiments, YbCCCa and YbCCAl. Despite the small size of the molecules, the electronic properties of each of the metal centers remain quite independent. This allows for photon cycling that is localized on each metal, exploiting their different advantages. Although these two molecules do not posses all of the desired properties for precision measurement experiments that specifically utilize a Yb atom, they suggest a more general class of promising molecules which contain Ca and a heavy metal that does not make use of photon cycling for precision measurement. This general recipe for hypermetallic small molecules likely allows for the laser cooling of a wide variety of heavy metal atoms via a Ca center, which is one potential path towards ultra precise next generation experiments.

\begin{acknowledgements}
MJO acknowledges support from a US National Science Foundation Graduate Research Fellowship under Grant No. DEG-1745301, as well as useful conversations with Narbe Mardirossian, Alec White, Ivan Kozyryev, Zhendong Li, and Garnet Chan. NRH acknowledges funding from the Heising-Simons Foundation and the NIST Precision Measurement Grants Program, as well as useful conversations with Tim Steimle, Wes Campbell, Eric Hudson, Svetlana Kotochigova, Maxim Ivanov, and Anna Krylov.  NRH also acknowledges the ``Molecules Functionalized with Optical Cycling Centers'' collaboration, which is supported by the U.S. Department of Energy (Award DE-SC0019245).

\end{acknowledgements}

\appendix

\section{Magnitude of spin-orbit coupling effects in YbCCCa}
\label{sec:SOC}

The results in Table \ref{tab:trueLevelsYbCCCa} show some spin-orbit sub-levels of the $A$ states which are heavily mixed between $^3\Pi$ and $^1\Pi$ configurations, along with some sub-levels which are purely triplets. The mixed sub-levels arise from linear combinations of the $m_s=0$ sectors in the singlet and triplet manifolds. No such linear combinations between singlet and triplet manifolds can be made for the $m_s=\pm 1$ triplet sectors (due to symmetry), so there are additional sub-levels of the $A$ states which are purely triplet. Since the energy gap between the mixed sub-levels and the pure sub-levels is only $\sim 100$ MHz, the radiative decay lifetime between the pure sub-levels and the mixed ones is essentially infinite relative to experimental timescales.

For a detailed understanding of why the $m_s=0$ sectors mix so strongly, we can perform an explicit analysis on a small part of the spin-orbit matrix $H_{\mathrm{SOC}} = \sum_i \alpha_i \vec{L}_i \cdot \vec{S}_i$, where $\vec{L}_i$ is the orbital angular momentum, $\vec{S}_i$ is the electron spin, and $\alpha_i$ is a spin-orbit constant. We will examine here only the Ca-centered $A$ state, although identical reasoning extends to the mixing of the Yb-centered sub-levels as well. In the spin-pure basis $\{A^1\Pi_{\mathrm{Ca}}, A^3\Pi_{\mathrm{Ca}}\}$ (and in units of cm$^{-1}$), we computed,

\begin{eqnarray}
H_0 & = &
\begin{bmatrix}
-418,852,888 & 0 \\
0 & -418,852,890
\end{bmatrix} \\
H_{\mathrm{SOC}} & = & 
\begin{bmatrix}
15198.43 & 20.74i \\
-20.74i & 15196.52
\end{bmatrix}.
\end{eqnarray}
Note that the off-diagonal terms are of the same order as the spin-orbit constant in CaOH~\cite{Kovacs1958,Bernath1985}. Thus the ``full" Hamiltonian is given by $H_{\mathrm{tot}} = H_0 + H_{\mathrm{SOC}}$,
\begin{equation}
H_{\mathrm{tot}} = 
\begin{bmatrix}
-418837689.57 & 20.74i \\
-20.74i & -418837693.48
\end{bmatrix}.
\end{equation}

This matrix can be analyzed by a simplified matrix of the form,
\begin{equation}
    \tilde{H}_{\mathrm{tot}} =
\begin{bmatrix}
N + \epsilon & \chi \\
-\chi & N-\epsilon
\end{bmatrix},
\end{equation}
where $2\epsilon$ is the splitting between between $A^1\Pi$ and $A^3\Pi$ due to standard correlation effects, while the mixing $\chi$ is due to spin-orbit effects. Note that the very large energies on the diagonals of $H_0$ and $H_{\mathrm{SOC}}$ mostly come from the quantum chemical ``background'' (ie. the core electrons), which is computed to very high accuracy. Typically only small components of the valence energy are subject to significant possible errors, so error in the splitting $\epsilon$ should be considered as a percentage of $\epsilon$, not a percentage of $N$.   

In the limit $\epsilon\to0, |\chi|>0$, the eigenvectors of $\tilde{H}_{\mathrm{tot}}$ approach the fully mixed $ \left[ i/\sqrt{2},\; \pm 1/\sqrt{2} \right]^T$, while in the limit $\chi\to0, |\epsilon|>0$, the eigenvectors approach the completely unmixed $ \left[ 1,\;  0 \right]^T$, $ \left[ 0,\;  1 \right]^T$. In the case of $H_{\mathrm{tot}}$, we have $\epsilon \sim 2$ cm$^{-1}$ and $\chi \sim 20i$ cm$^{-1}$, which gives highly mixed eigenvectors: $\left[ 0.673i,\; -0.740 \right]^T$, $\left[ 0.740i,\;  0.673 \right]^T$.

Thus, the strong mixing effect emerges because the singlet-triplet energy gap is small compared to the magnitude of the spin-orbit coupling term.  However, as we just discussed, $\epsilon$ and $\chi$ are the parts of the computation which are highly sensitive to the electronic structure and are subject to possible errors based on the computational methodology. Nonetheless, we expect our conclusion of ``strong spin-orbit mixing'' to be valid because even if we assume our calculated values of $\epsilon$ and $\chi$ both have very large errors of $\sim 90 \%$ each, we still have $|\epsilon/\chi| \approx 1$. Using this ratio to compute the eigenvectors of $\tilde{H}_{\mathrm{tot}}$, we see that they are still quite mixed:
$\left[ 0.924i,\; 0.383 \right]^T$, $\left[ -0.383i,\;  0.924 \right]^T$.

Furthermore, in the context of potential experiments the spin-orbit mixing is only negligible in the limit when $|\chi/\epsilon| \approx 10^{-4}$. Assuming $\chi$ is approximately correct due to its similarity with the CaOH results, this limit could only be reached if the error in $\epsilon$ is $10^5 - 10^6 \%$, which we deem unlikely based on our computational methodology (discussed in Section \ref{sec:compDetails}).

\bibliography{main}
\end{document}